# Supersonic Turbulent Flows and the Fragmentation of a Cold Medium

Paolo Padoan

*Theoretical Astrophysics Center*
*Blegdamsvej 17, DK-2100 København ø, Denmark*
*and*
*Dept. of Astronomy - University of Padova*
*Vicolo dell'Osservatorio 5, I-35122 Padova, Italy*

**ABSTRACT**

The role played by velocity fields in the fragmentation of a cold medium and in the formation of protostars is studied.

The velocity field is modeled with a compressible turbulent flow. A supersonic turbulent velocity field can fragment the medium into clumps of mass smaller than a local Jeans' mass, and therefore stabilize the medium against the formation of protostars. Based on this idea, the protostar formation efficiency and the protostar mass distribution are determined as functions of the following ambient parameters: average density $n_0$, average temperature $T_0$, r.m.s. turbulent velocity $\sigma_{v,0}$ (or its Mach number $\mathcal{M}_t$), postshock cooling time (e.g. chemistry).

The main results are:

(i) the protostars mass distribution and its dependence on the ambient parameters are quantified;

(ii) the characteristic protostar mass is $M_{J,cl} \propto n_0^{-1/2} T_0^2 \sigma_{v,0}^{-1}$;

(iii) the protostar formation efficiency $e$ is higher for larger mean density, larger mean temperature, lower velocity dispersion on a given scale and longer postshock cooling time (e.g. lower metallicity): $e \propto n_0^{\frac{3}{2}(\beta-1)} T_0^{\beta-1} \sigma_{v,0}^{-5(\beta-1)} L_0^{3(\beta-1)}$, where $\beta > 1$ is the exponent of the clump mass distribution;

(iv) the efficiency is quite sensitive to the ambient parameters and therefore to the dynamical evolution of the star forming system.

## 1 INTRODUCTION

### 1.1 Star formation in galaxies

The observational properties of galaxies, upon which cosmological models rely, cannot be interpreted without a knowledge of their star formation history. Examples of such properties are integrated colors, surface brightness and chemical abundances.

Also on purely theoretical grounds star formation is an essential part of the theory of galaxies formation, since the dissipative collapse of a protogalaxy is eventually arrested by its fragmentation into stars. Star formation is therefore a fundamental cosmological problem.

The study of the evolution of a single protostar, while being a necessary part of the theory of star formation, is not the key point in the context of galaxy formation. Rather it is the statistical description of the fragmentation of a large scale environment with given initial and boundary conditions. Because of the complex network of interacting physical phenomena in (proto)galaxies large scale protostars formation is certainly a chaotic process, so that its description must be statistical.

Protostar formation in galaxies is normally described with simple parametrizations of the star formation efficiency such as Schmidt's law (Schmidt 1959; see Kennicutt 1989) or more complex (Wyse and Silk 1989; Dopita and Ryder 1994), and often based on the idea of self regulated star formation (McKee and Lin 1988) and disk instabilities (see the review by Elmegreen 1993). Feedback arguments are also often used (e.g. Whitworth 1979, Lin and Murray 1992).

The problem is that of predicting protostar formation efficiency (or rates) and protostellar masses in a large scale environment with given dynamics and chemistry and which is initially devoid of stars. Without a statistical description of the formation of protostars our knowledge on protostar evolution will not have the desired impact upon the theory of galaxy formation and evolution.

The present work is an attempt to predict protostar formation efficiency and protostellar masses as functions of ambient properties such as mean density, mean temperature, turbulent velocity dispersion on a given scale, chemistry (postshock cooling time) by focusing only on one aspect of the process of star formation, namely the hydrodynamics of the fragmentation of a cold medium when high Reynolds and Mach numbers flows are present.

### 1.2 Supersonic turbulence and statistical description of protostars formation

Many astrophysical environments possess highly supersonic flows with formation of shock waves.

When the Reynolds number of the flow, that measures the ratio of the inertial term, $\mathbf{v} \cdot \nabla \mathbf{v}$, relative to the viscous term, $\nu \nabla^2 \mathbf{v}$, in the Navier-Stokes equation, is very large an important instability occurs: a regular flow (normally referred to as laminar) is not possible, because the slightest disturbance will initiate the transfer of energy between different scales in the flow, and the molecular viscosity cannot compete with this energy transfer due to the inertial term (see Batchelor 1967). The energy transfer occurs in the form of energy cascades, normally from the large scales to the small ones.

The manifestation of this instability is generically called turbulence (see Batchelor 1967, Landau and Lifshitz 1987, Frisch and Orszag 1990).

The importance of the onset of turbulence is that it allows for the use of equilibrium laws in the statistical description of the flow on small scales (see Batchelor 1953, Monin and Yaglom 1971, Tennekes and Lumley 1972). Therefore while some astrophysical flows are certainly chaotic in the sense of not being predictable, others are chaotic and yet predictable in a statistical sense because they are unstable to turbulence.

If this is the case for some astrophysical supersonic flows it is possible that the formation of shocks in them can be described as the effect of turbulence and therefore predicted in a statistical sense. Note that it was recognized long ago that the Reynolds number in the interstellar matter on the scale of $100 pc$ has a very high value of the order of $10^6$ (von Weizsker 1951).

The problem is particularly interesting in the star forming environments. Understanding protostar formation is mainly a matter of predicting the local occurrence of gravitational instabilities in large regions of the interstellar gas, not necessarily unstable as a whole. The simplest way to do this is to assume that the gas fragments into protostars whose mass equals the Jeans' mass in the large region which is supposed to be homogeneous and with little velocity fluctuations (Jeans 1902, Hoyle 1953, Fowler and Hoyle 1963).

This approach is not even a first approximation if the interstellar gas is organized in supersonic chaotic flows because of the large velocity and density gradients.

Nevertheless if the motions are in the regime of fully developed turbulence it could be possible, as mentioned above, to predict the formation of shocks using the statistical theory of turbulence and apply simple instability arguments to the shocked regions.

In this way the protostar formation will be shown to depend not only on the mean density and temperature of the environment, but also on the turbulent velocity dispersion responsible for the compressions in the gas, and on the chemistry responsible for the postshock cooling. Clearly the more that is known about the turbulent model flow, the more can be inferred about the shock distribution and the protostar formation.

This simple way of describing protostar formation is certainly not comprehensive of all the physics involved in the problem of large scale star formation (especially because magnetic fields will not be considered in the following). Nevertheless it is interesting to study the role of hydrodynamical supersonic turbulence in the formation of protostars, at least in environments where the magnetic field is known not to play a dominant part. [*]

Larson (1981) suggested that present day star formation in the Galaxy could occur in shocks generated by the supersonic turbulence that could be the reason for the width of the molecular lines emitted by the interstellar gas.

Hunter (1979) and Hunter and Fleck (1982) have studied the influence of the velocity field upon star formation, but without taking into account the formation of shocks.

Léorat et al. (1990) have shown, with two-dimensional numerical simulations, that compressible turbulence can affect the gravitational instability in the ISM.

Elmegreen (1993) has studied the gravitational instability of shocked regions of the ISM and shown how the clumpy structure of molecular clouds and their star formation mechanism can be controlled by the turbulent Mach number.

### 1.3 Overview of the paper

Section 2 contains a discussion of the physical assumptions used throughout the work.

The fragmentation controlled by the turbulent velocity field is described in terms of postshock high density clumps

---

[*] Chandrasekhar (1958) suggested a correction of the Jeans' mass due to turbulent pressure; Arny (1971) used the idea of turbulent pressure to predict the protostar mass function; Bonazzola et al. (1987) studied the influence of the turbulent spectrum on the gravitational instability, restudying the effect of turbulent pressure. That turbulence acts as a pressure source against gravity has been shown by Bonazzola et al. (1992) (see also the simulations by Léorat et al. 1990).

Nevertheless, in order to be the dominant source of pressure the turbulent motions must be supersonic (only in this case their role is significant in the cited works); if this is the case and if the medium has a short cooling time (in the sense specified in the paper) strong shock formation leading to very high density contrasts is likely to be the main fragmentation mechanism due to turbulence, and turbulent pressure is of secondary importance.

in section 3, where the properties and stability of the clumps are studied.

The concept of turbulent fragmentation is applied to protostar formation in section 4; in particular formulae for the mass distribution and the formation efficiency of protostars are given.

Section 5 is a discussion of the role of turbulent fragmentation in controlling the formation of protostars.

The work is summarized in section 6.

A list of the symbols used throughout the paper is given in the appendix.

## 2 PHYSICAL ASSUMPTIONS

A number of physical assumptions will be used in order to study the fragmentation controlled by a turbulent velocity field.

The first is the exclusion of the magnetic field. The observational data on the intensity of the magnetic field in dense clouds (Heiles and Stevens 1986; Crutcher, Kazès and Troland 1987; Kazès et al. 1988; Crutcher et al. 1993) would not justify this assumption if the present Galactic ISM were the subject of the study; nevertheless it is important to understand the role played by purely hydrodynamical processes in the fragmentation of a cold medium. The whole hydromagnetic treatment of the problem is under study through numerical simulations and will be presented in a later paper (Nordlund and Padoan, in preparation).

It will be shown that if a protostar is identified with one Jeans' mass in the fragmented gas, the protostars mass distribution and formation rates can be determined once the energy spectrum of turbulence and the mass and density distribution of postshock clumps are specified. Note that the essential ingredient is the energy spectrum of turbulence, since this fixes the typical density and mass of the clumps and the typical protostellar mass. It is therefore assumed that the ISM flows are in the regime of fully developed turbulence, as their huge Reynolds number indicates (Von Weiszker 1951; Scalo 1987), so that a power law velocity spectrum can be assumed.

### 2.1 The picture

A complex system of isothermal shock waves arises in the ISM due to the presence of a turbulent and highly supersonic velocity field. After one dynamical time of the largest scale the ISM matter distribution is very clumpy, with density contrasts of the order of the mean squared Mach number of the velocity fluctuations. For the sake of definitiveness it will be assumed in the following that the mass distribution can be described as an ensemble of clumps that are formed after shock waves. A typical mass scale of turbulent fragmentation will be called "typical clump" and a clump mass distribution will be defined.

Some clumps happen to be dense and massive enough to undergo gravitational instability and fragment into protostars. Clearly this way of forming protostars is highly dependent on the density and on the mass of the clumps. More precisely, since in typical ISM conditions most clumps will be stable against gravitational collapse, it is the intermittency in the statistics of the density field (shaped by supersonic turbulence) that determines what fraction of the total mass becomes unstable and is converted into protostars.

### 2.2 Turbulent spectrum

Supersonic velocity dispersions $\sigma_v$ are observed on scale $r$, according to the Larson's relation (Larson 1981):

$$\sigma_v \approx 1 km s^{-1} \left(\frac{r}{1pc}\right)^{0.4} \qquad (1)$$

valid in the range of scales $0.1 - 100 pc$ (Larson 1981). By including HI regions this relation is valid up to $1000 pc$ (Larson 1979). The correlation between line width and linear size in the molecular ISM has been confirmed (with slightly different exponents and normalization factors) by a number of authors (Leung, Kutner and Mead 1982; Myers 1983; Quiroga 1983; Sanders, Scoville and Solomon 1985; Arquilla and Goldsmith 1985; Dame et al. 1986; Falgarone and Pérault 1987; Fuller and Myers 1992).

Despite the numerous interpretations of this relation in terms of a Kolmogorov turbulent cascade (Kolmogorov 1941), it should be remembered that Larson's relation is not a direct measure of the velocity power spectrum, so that the observed exponent, $0.4 - 0.5$, could arise from different mechanisms rather than from the presence of a turbulent spectrum such as the one expected from models of turbulence.

On the other hand a number of authors have tried to measure autocorrelation and structure functions of ISM velocity fields directly (Scalo 1984; Dickman and Kleiner 1985; Kleiner and Dickman 1985; Kleiner and Dickman 1987; Hobson 1992, Miesch and Bally 1993). No definitive conclusion as to which model of turbulence is most appropriate to describe the ISM has been reached.

When modeling the ISM turbulence one has therefore to rely on the statistical description of numerical turbulence. Direct numerical simulations of highly supersonic turbulence are still very difficult, but there are good indications that an energy spectrum:

$$E(k) \propto k^{-2} \qquad (2)$$

is to be expected for the compressional modes (Passot, Pouquet and Woodward 1988; Porter, Pouquet and Woodward 1992) that are responsible for the fragmentation (formation of shocks). Such a spectrum is also to be expected for a flow made up of discontinuities (Passot and Pouquet 1987), and it is known to arise from Burgers' equation (Burgers 1940, 1974; Gotoh and Kraichnan 1993).

Given these indications, the power spectrum (2) will be assumed.

In equation (2) $E(k)$ is the integration, over a spherical shell of radius $k$ (modulus of the wave vector), of the trace $\Phi_{ii}(\mathbf{k})$ of the spectrum tensor $\Phi_{ij}(\mathbf{k})$ which is the Fourier transform of the correlation tensor $R_{ij}(\mathbf{r})$:

$$R_{ij}(\mathbf{r}) \equiv \overline{u_i(\mathbf{x},t)u_j(\mathbf{x}+\mathbf{r},t)} \qquad (3)$$

where $u_i$ is the flow velocity.

The trace $\Phi_{ii}(\mathbf{k})$ represents the kinetic energy per unit mass at wave vector $\mathbf{k}$, as can be seen from:

$$R_{ii}(0) = \overline{u_i u_i} = \int\int\int_{-\infty}^{\infty} \Phi_{ii}(\mathbf{k})d\mathbf{k} \qquad (4)$$

Thus one can write:

$$\overline{u_i u_i} \propto \int_k^\infty E(k')dk' \propto k^{-1} \propto r \tag{5}$$

Therefore if the spectrum (2) is used, the r.m.s. velocity $\sigma_v(r)$ on the scale $r$ (or the velocity of "eddies" of size $r$) is:

$$\sigma_v(r) \propto r^{1/2} \tag{6}$$

Notice that this relation means that the dynamical time, $r/\sigma_v(r)$, increases with the linear scale according to:

$$\tau_{dyn}(r) \propto r^{1/2} \tag{7}$$

### 2.3 Clumps density and mass distributions

The shock waves in the flow result in very non-Gaussian statistics for the velocity gradient. The statistics of divergence of the velocity field has been shown to be extremely intermittent in numerical simulations of compressible turbulence (Lee, Lele and Moin 1991). The density distribution is thus expected to be non-Gaussian; Vázquez-Semadeni (1994) has found that the density field in two-dimensional simulations of compressible turbulence is lognormally distributed; this is confirmed in preliminary results of three-dimensional numerical simulations of highly supersonic turbulence (Nordlund and Padoan, in preparation).

Therefore the density distribution of the clumps into which the postshock gas is fragmented is assumed to be lognormal:

$$P(x)dx = \frac{x^{-1}}{(2\pi)^{1/2}} \exp\left(-\frac{1}{2}\ln^2 x\right) dx \tag{8}$$

where $x = n/n_{cl}$, with $n$ number density and $n_{cl}$ the density of the typical clumps (defined by equation (10) that depends on the normalization of the turbulent spectrum (see next section).

It will also be assumed that the density and mass distributions are statistically independent so that the joint probability can be written as a product of the two distributions (this will simplify the derivation of the protostars mass distribution). This is justified by the fact that the density is mainly determined by intense shocks on large scales while the fragmentation into small clumps is controlled by the less intense shocks on small scales.

The clumps into which the dense postshock gas is fragmented are characterized by a typical mass, $M_{cl}$ (see equation (15), that depends on the turbulent spectrum (see next section); nevertheless it is clear that larger clumps can exist, even if less numerous. For simplicity the mass distribution is assumed to be a power law:

$$P_{cl}(m_{cl})dm_{cl} = (1-\beta)m_{cl}^{-\beta}dm_{cl} \tag{9}$$

for $m_{cl} \geq 1$, where $m_{cl}$ is the mass in units of the typical clump mass, $M_{cl}$, defined in the next section, and $\beta > 1$.

Note that this mass distribution (as all the following ones) is defined in such a way that the number of objects per mass interval is $N(m)dm \propto \frac{P(m)}{m}dm$, that is $P(m)$ is proportional to the number of objects per logarithmic mass interval.

Several observations yield the value $\beta \approx 0.6$, for both giant molecular clouds and clouds cores. The subsection 5.1 explains why there is no contradiction between the observations and the assumed clump mass distribution.

## 3 TURBULENT FRAGMENTATION

### 3.1 Clumps properties

Given the turbulent velocity (6), it is possible to define the transition scale $r_{tr}$ below which the velocity becomes subsonic:

$$\mathcal{M}(r_{tr}) \equiv 1 \tag{10}$$

where $\mathcal{M}(r)$ is the Mach number of the turbulent velocity dispersion on the scale $r$, relative to the isothermal sound speed of the medium.

The scale $r_{tr}$ is the typical scale of the fragmentation into clumps due to a turbulent velocity field. In a cloud of mass $M_{cloud}$ it corresponds to a mass given by:

$$M_{tr} \equiv \frac{M_{cloud}}{\mathcal{M}_t^6} \tag{11}$$

where $\mathcal{M}_t$ is the Mach number of the cloud velocity dispersion; $M_{tr}$ is between 0.01 and $10 M_\odot$ in typical giant molecular clouds of the galactic disc. This is an interesting range of masses for protostar formation.

If the postshock cooling time is supposed to be shorter than the dynamical time of the shocks (see below), the postshock density on the scale $r_{tr}$ is estimated by the isothermal jump condition:

$$\frac{\rho_2}{\rho_1} \approx \mathcal{M}_t^2 \equiv \left(\frac{\sigma_{v,0}}{c_S}\right)^2 \tag{12}$$

where $\rho_1$ and $\rho_2$ are the gas volume density before and after the shock, $c_s$ is the isothermal sound speed and $\mathcal{M}_t$ is the Mach number of the ambient r.m.s. turbulent velocity, $\sigma_{v,0}$, on the scale $L_0$.

The reason why this large scale Mach number is used is that the stability of the shocked structures formed after one dynamical time of the large scales will be later studied (this will be used to define the efficiency of the formation of protostars). In fact most of the density contrast relative to the mean density is produced by large scale shocks after one dynamical time of the large scales. Shocks on small scales are less intense (see equation (6)) and their role is mainly that of fragmenting the gas into smaller subunits, since their contribution to the density enhancement over the mean ambient density is irrelevant.

The relations (6), (10) and (12) determine the mass and the density of the typical postshock clumps into which the medium is fragmented. The typical mass $M_{cl}$ is:

$$M_{cl} = \frac{4\pi}{3}\rho_0 r_{tr}^3 \tag{13}$$

where $\rho_0$ is the ambient average density and the transition scale defined by (10) is:

$$r_{tr} = 1pc \left(\frac{\mathcal{M}_t}{10}\right)^{-2} \frac{L_0}{100pc} \tag{14}$$

where $L_0$ is the linear size of the region whose r.m.s. turbulent velocity is $\sigma_{v,0}$ (e.g. the size of the energy containing turbulent eddies).

Therefore:

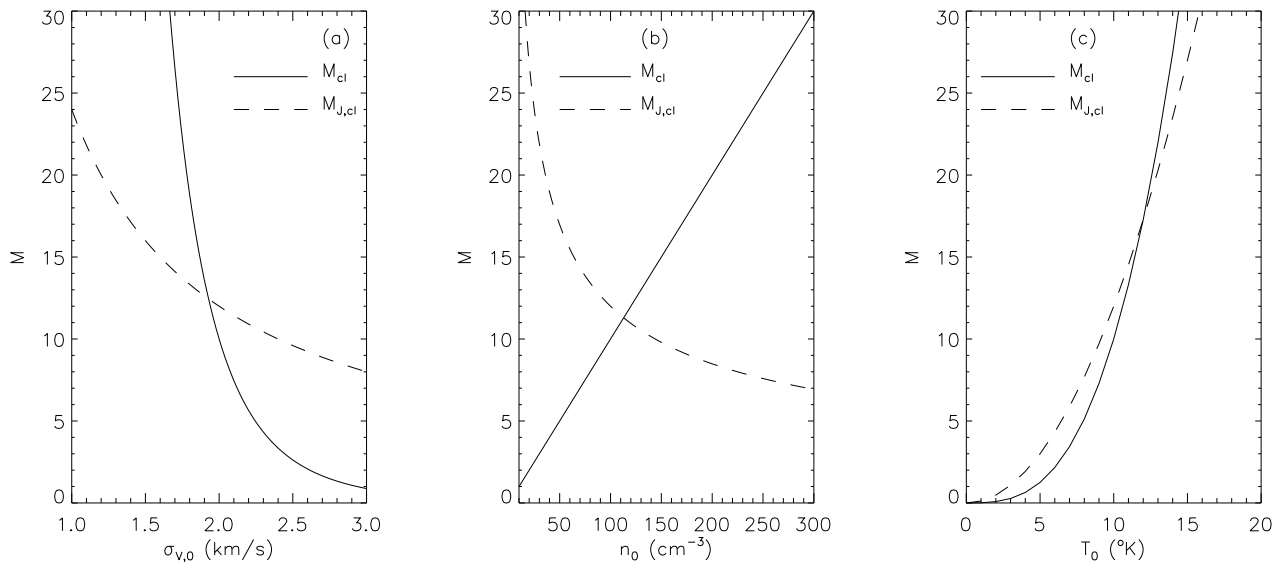

**Figure 1.** The typical mass of the postshock clumps, $M_{cl}$, and the Jeans' mass in the clumps, $M_{J,cl}$, are plotted versus the ambient r.m.s. turbulent velocity (a), the mean density (b) and the mean temperature (c). In (a) $n_0 = 100 cm^{-3}$ and $T_0 = 10°K$, in (b) $\sigma_{v,0} = 2 km/s$ and $T_0 = 10°K$, in (c) $\sigma_{v,0} = 2 km/s$ and $n_0 = 100 cm^{-3}$ have been used. In all three plots $L_0 = 100pc$. The mass is expressed in solar masses.

$$M_{cl} \approx 10 M_\odot \left(\frac{n_0}{100 cm^{-3}}\right)\left(\frac{\sigma_{v,0}}{2km/s}\right)^{-6}\left(\frac{T_0}{10K}\right)^3\left(\frac{L_0}{100pc}\right)^3$$
$$\approx 10 M_\odot \left(\frac{n_0}{100 cm^{-3}}\right)\left(\frac{\mathcal{M}_t}{10}\right)^{-6}\left(\frac{L_0}{100pc}\right)^3 \quad (15)$$

where $n_0$ is the number density corresponding to $\rho_0$.

From the relation (12) the number density in the clump is:

$$n_{cl} \approx n_0 \mathcal{M}_t^2 \quad (16)$$

If the postshock cooling time, $\tau_{cool}$, is longer than the shock dynamical time, $\tau_{dyn}$, (the time required by the shock to traverse the clump's mass) no postshock structure of high density contrast of mass comparable to the one traversed by the shock can be formed. The radiative postshock zone would extend through the whole shocked gas, and the gas would simply disperse, since the energy of compression is expected to be larger than the gravitational energy on this scales (see Larson 1988). Because of the turbulent energy spectrum $k^{-2}$ (equation (2)) this timing condition for the shocks to be radiative can easily be translated into a minimum mass, $M_{rad}$, for postshock clump formation:

$$M_{rad} \approx M_0 \left(\frac{\tau_{cool}}{\tau_{dyn,0}}\right)^6 \quad (17)$$

where $M_0$ is the mass contained in the large scale with dynamical time $\tau_{dyn,0}$.

Since for a large system $\tau_{cool} < \tau_{dyn,0}$ (see equation (7)) and because of the large exponent in the relation (17) $M_{rad} \ll M_0$ is generally satisfied.

The typical mass of the fragments produced by the turbulent velocity field is therefore $max(M_{cl}, M_{rad})$. For a solar metallicity the cooling time in molecular clouds is short enough that $M_{rad} < M_{cl}$. The low metallicity regime with $M_{rad} > M_{cl}$ is of great importance to describe population II star formation. There are a number of interesting issues relating to this regime and this will be dealt with in a later paper.

### 3.2 Clumps stability

#### 3.2.1 Turbulent pressure

Molecular clouds have too low temperatures $(10 - 50K)$ to be supported by their thermal pressure (Larson 1981; Bonazzola et al. 1987, 1992; Leorat, Passot and Pouquet 1990). The ISM turbulence has been invoked as an alternative to the magnetic field (see Mouschovias 1987) as the pressure source against the gravitational collapse of molecular clouds and different mechanisms of energy injection into the turbulent cascade have been proposed (von Weizsaecker 1951; Fleck 1980,1981; Norman and Silk 1980; Scalo and Pumphrey 1982; Franco and Cox 1983; Henriksen and Turner 1984; McCray and Kafatos 1987). A possible solution to the problem of the short dissipation time of the turbulent motions compared to the estimated lifetime of giant molecular clouds (Blitz and Shu 1980), is that the dissipation of turbulence is reduced as a consequence of the fragmentation of the clouds into clumps (Scalo and Pumphrey 1982).

Turbulent pressure could be important also inside the postshock clumps into which the medium is fragmented by supersonic turbulence (see subsection 3.1), but it is probably not the case because of the following arguments.

The density contrast between the clumps and the surrounding medium and the resultant low volume filling fraction of the clumps is likely to interrupt the hydrodynamic turbulent cascade of energy from the large scales. Moreover the fragmentation inside the clumps cannot be invoked to make the dissipation time longer, since for the definition

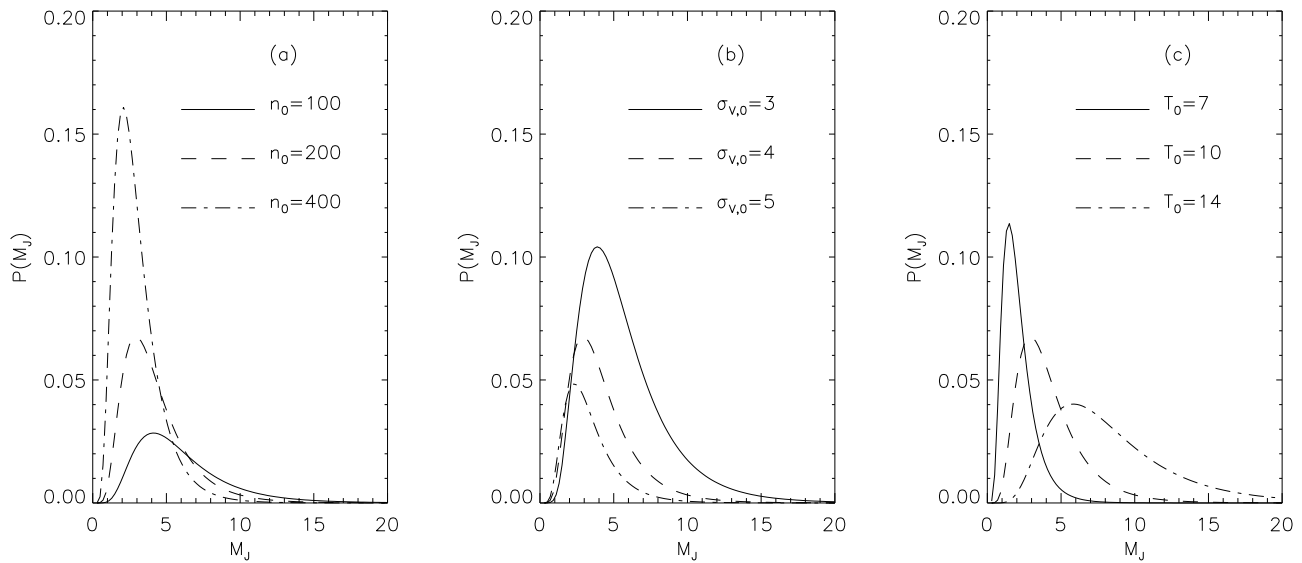

**Figure 2.** Protostars mass distribution from equation (18). The function is plotted for different values of the ambient density (in $cm^{-3}$) (a), turbulent velocity dispersion (in $km/s$) (b) and temperature (in $K$) (c). In (a) $T_0 = 10K$ and $\sigma_{v,0} = 4km/s$; in (b) $n_0 = 200cm^{-3}$ and $T_0 = 10K$; in (c) $n_0 = 200cm^{-3}$ and $\sigma_{v,0} = 4km/s$. The mass is in solar masses. The clumps mass distribution has been modeled with a power law (see text) with exponent $\beta = 1.5$

(10) of the clumps no further fragmentation will occur inside them. Thus the turbulence inside the clumps is probably dissipated in a short time (of the order of, or less than, the dynamical time in the clumps). Finally turbulent pressure is not dominant over thermal pressure (as is obvious from the definition (10)), despite some increase of turbulent energy in the postshock gas is expected (see Rotman 1991, Jacquin, Cambon and Blin 1993). For these reasons the internal turbulent pressure in the clumps will not be used in the definition of the Jeans' mass inside the clumps.

### 3.2.2 Jeans' mass in the clumps

The Jeans' mass inside the clumps is:

$$\begin{aligned} M_{J,cl} &\approx 12 M_\odot \left(\tfrac{T}{10K}\right)^{3/2} \left(\tfrac{n_0}{100cm^{-3}}\right)^{-1/2} \left(\tfrac{n_{cl}}{100n_0}\right)^{-1/2} \\ &\approx 12 M_\odot \left(\tfrac{T}{10K}\right)^{3/2} \left(\tfrac{n_0}{100cm^{-3}}\right)^{-1/2} \left(\tfrac{\mathcal{M}_t}{10}\right)^{-1} \\ &\approx 12 M_\odot \left(\tfrac{T}{10K}\right)^{2} \left(\tfrac{n_0}{100cm^{-3}}\right)^{-1/2} \left(\tfrac{\sigma_{v,0}}{2km/s}\right)^{-1} \end{aligned} \quad (18)$$

where equation (16) has been used in the second expression and equation (12) (the definition of the Mach number) in the third, with the result of changing the temperature dependence and of introducing the ambient velocity dispersion. This Jeans' mass decreases with increasing ambient velocity dispersion (denser clumps are produced according to equation (16)) contrary to the case of internal turbulent pressure (see references in the footnote in section 1). This is in fact equivalent to the Jeans' mass at constant external pressure, if turbulent ram pressure is considered.

To study the stability of the typical clumps formed by the supersonic velocity field, equation (18) has to be compared with equation (15). This is illustrated in figs. 1.

As shown in figs.1 the mass of the typical clumps $M_{cl}$ becomes larger than the largest stable mass, $M_{J,cl}$, and therefore the clumps are unstable, for decreasing ambient r.m.s. turbulent velocity and for increasing mean density and temperature. The increase in $\sigma_{v,0}$ produces denser clumps (higher $\mathcal{M}_t$), but since their typical mass is also reduced, they are more stable.

If the postshock cooling time is such that $M_{rad} > M_{cl}$ (see eq. (17)), the mass to be compared with $M_{J,cl}$ is $M_{rad}$ and not $M_{cl}$. In this regime a long postshock cooling time means more massive and more unstable clumps than in the case of a shorter cooling time. Inefficient cooling acts against the fragmentation by the velocity field that otherwise tends to stabilize the medium against the gravitational instability by producing clumps smaller than the local Jeans' mass.

## 4 PROTOSTAR FORMATION

### 4.1 Protostar mass distribution

It is assumed that any unstable clump fragments into protostars whose mass equals the Jeans' mass inside the clump.[†] Since a density and a mass distributions of clumps are assumed to exist, with typical values $n_{cl}$ and $M_{cl}$ determined by the turbulent spectrum, a protostar mass distribution will also exist around the value $M_{J,cl}$.

If $n_{cl}$ in equation (18) is changed into $n$, that is the local number density and obeys the statistics (8), equation (18)

---

[†] Thin slabs are more stable than the simple Jeans' mass predicts (see for example Gilden 1984); nevertheless in this work it is assumed that the velocity field fragments the medium down to the small scales where the typical shocks are not very intense and therefore the postshock gas cannot be described as a thin slab.

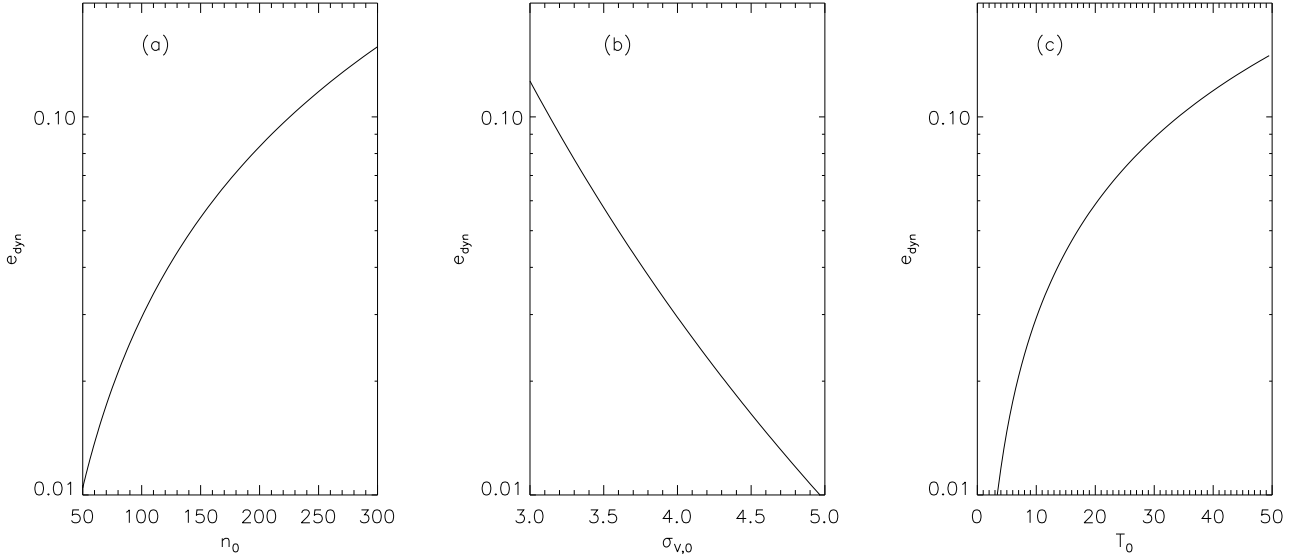

**Figure 3.** The protostar formation efficiency per ambient dynamical time is plotted versus the ambient density expressed in $cm^{-3}$ (a), the turbulent velocity dispersion in $km/s$ (b) and the ambient temperature in $K$ (c). In (a) $\sigma_{v,0} = 4 km/s$ and $T_0 = 10K$; in (b) $n_0 = 100 cm^{-3}$ and $T_0 = 10K$; in (c) $n_0 = 100 cm^{-3}$ and $\sigma_{v,0} = 4 km/s$. $L_0 = 100 pc$ and $\beta = 2.0$ in all three plots. The curves are power laws (see text).

gives the mass $M_J$ of the protostars formed in the clumps of density $n$.

The protostars mass distribution $P(m)$, with $m = M_J/M_{J,cl}$, is therefore determined by:

$$P(m) = P(x)\frac{dx}{dm} \qquad (19)$$

($x = n/n_{cl}$) where $P(x)$ is from equation (8) and $dx/dm$ comes from equation (18) after the substitutions $M_{J,cl} = M_J$ and $n_{cl} = n$. The protostar mass distribution is thus:

$$P(m)dm = \frac{2m^{-1}}{(2\pi)^{1/2}} \exp(-2\ln^2 m) dm \qquad (20)$$

If the mass distribution of clumps is $P_{cl}(m_{cl})$, the protostar mass distribution becomes:

$$P(m)dm = P(m)dm \int_{m\frac{M_{J,cl}}{M_{cl}}}^{\infty} P_{cl}(m_{cl})dm_{cl} \qquad (21)$$

which means that protostars of mass $M_J$ can be formed only in clumps of mass $> M_J = m M_{J,cl}/M_{cl}$.

The factorization into the product of the protostar mass distribution and the integral of the clumps mass distribution is allowed by the assumption of statistical independence of the density and mass distributions of the clumps (see subsection 2.2).

Using the clumps mass distribution (9), the protostar mass distribution, equation (21), is further transformed into:

$$P(m)dm = \frac{2m^{-1}}{(2\pi)^{1/2}} \exp(-2\ln^2 m) dm,$$
$$if\, m < M_{cl}/M_{J,cl};$$
$$P(m)dm = \frac{2m^{-\beta}}{(2\pi)^{1/2}}\left(\frac{M_{cl}}{M_{J,cl}}\right)^{\beta-1} \exp(-2\ln^2 m) dm,$$

$$if\, m \geq M_{cl}/M_{J,cl} \qquad (22)$$

Several examples of protostar mass distributions are plotted in fig.s 2, for different values of the ambient parameters.

In each of fig.s 2 three protostar mass distributions are plotted for different values of $n_0$, $\sigma_{v,0}$ and $T_0$ respectively. The function plotted is equation (22) with $\beta = 1.5$.

The protostar mass distribution has a single maximum corresponding approximately to $0.5 M_{J,cl}$, and goes to zero for both very large and very small protostellar masses.

It has to be emphasised that the characteristic mass of a protostar given by equation (18) differs from the traditional Jeans' mass in the temperature dependence and because it depends also on the ambient velocity dispersion $\sigma_{v,0}$:

$$M_{J,cl} \propto n_0^{-1/2} T_0^2 \sigma_{v,0}^{-1} \qquad (23)$$

This characteristic protostellar mass decreases with increasing velocity dispersion (because denser clumps are produced according to (16)), and is equivalent to the Jeans' mass at constant external pressure, if turbulent ram pressure is considered.

If the postshock cooling time is such that $M_{rad} > M_{cl}$ (see equation (17)) the protostar mass distribution is the same as in equation (22), but with the substitution $M_{cl} = M_{rad}$, because no high density contrast bound clump of mass $< M_{rad}$ can be formed, as explained in the end of the subsection 3.1.

### 4.2 Protostar formation efficiency

The protostar formation efficiency is simply determined by the mass fraction of unstable clumps relative to the total mass. The discussion about the postshock clumps stability presented in section 3 is therefore equivalent to the present discussion of the efficiency of protostar formation. The scenario is now complicated by the introduction of a mass distribution for the clumps and of a distribution for the Jeans' mass inside them (that are supposed to be statistically independent).

The mass distribution (21) expresses the fraction of the total mass contained in clumps whose Jeans' mass relative to $M_{J,cl}$ is between $m$ and $m + dm$ and whose mass is $> m$. Therefore the integration of (21) over $m$ gives the fraction of total mass contained in unstable clumps (or equivalently in protostars):

$$e_{dyn} = \frac{M_*}{M_{tot}} = \int_0^{m_{max}} P(m) dm \qquad (24)$$

The compression ratio $\propto \mathcal{M}_t^2$, used to determine the clumps density (see section 3), is experienced by the whole environment after one dynamical time of the large scale $L_0$:

$$\tau_{dyn,0} = \frac{L_0}{\sigma_{v,0}} \qquad (25)$$

where $\sigma_{v,0}$ is the r.m.s. velocity on the scale $L_0$. Therefore a time equal to $\tau_{dyn,0}$ is necessary to convert into protostars as much mass as expressed by equation (24).

For this reason the efficiency, defined as the protostar formation rate per unit mass, is:

$$e = \frac{e_{dyn}}{\tau_{dyn,0}} \qquad (26)$$

In figs. 3 the dependence of the efficiency on ambient parameters is shown: the efficiency per dynamical time $e_{dyn}$, equation (24), is plotted versus the mean density, the ambient r.m.s. turbulent velocity and the mean temperature respectively. The clumps mass distribution is modeled with a power law with exponent $\beta = 1.5$ and $M_{rad} < M_{cl}$ is supposed to be valid.

The efficiency grows with $n_0$ and $T_0$ and with decreasing $\sigma_{v,0}$, as the ratio $M_{cl}/M_{J,cl}$ (see subsection 3.2).

The curves in fig.3 follow the power laws:

$$e_{dyn} \propto n_0^{\frac{3}{2}(\beta-1)} \sigma_{v,0}^{-5(\beta-1)} T_0^{\beta-1} L_0^{3(\beta-1)} \qquad (27)$$

from $e_{dyn} \approx 0.01$ up to values as high as $e_{dyn} \approx 0.9$.

This can be also seen qualitatively since the factor that multiplies the protostar mass distribution equation (22) for $m \geq M_{cl}/M_{J,cl}$, $(M_{cl}/M_{J,cl})^{\beta-1}$, has exactly the same dependence on the ambient parameters as $e_{dyn}$ expressed in equation (27).

It is interesting that while it is possible that the efficiency per dynamical time $\approx 3\%$ at $T_0 = 10K$, the efficiency in the same environment warmed up to $T_0 = 30K$ increases to $\approx 9\%$ (for $\sigma_{v,0} = 4km/s$ on the scale $L_0 = 100pc$ and $n_0 = 100cm^{-3}$). This fact seems to indicate that wherever the mode of protostar formation here discussed is relevant, the protostar formation rate can be increased by the warming up of the ambient medium due to the first generation of stars (positive feedback). The increased temperature will also increase the protostars mass (fig. 2.c) with the effect of an even more efficient warming of the environment; this makes the feedback stronger (and the characteristic protostar mass increase with time).

A negative feedback could instead result from the injection of mechanical energy due to the products of protostar formation, if this could increase the $\sigma_{v,0}$ (see fig. 3.b).

If the ambient chemistry is such that $M_{rad} > M_{cl}$ (see equation (17)) $M_{rad}$ should be used instead of $M_{cl}$ for estimating the efficiency; if $M_{rad}/M_{cl} \gg 1$ the increase in the efficiency could be drastic.

## 5 DISCUSSION

It is a common assumption in theories of galaxy formation that, when the cooling time of the gas is shorter than the free fall time of the protogalaxy, fragmentation into stars occurs (Hoyle 1953; White and Rees 1978). This notion is based on linear gravitational instability and its applicability to a flow with certainly nonlinear velocity and density fluctuations is dubious.

Since the flow is turbulent and supersonic the study of the fragmentation of the protogalactic gas when cool and with short cooling time is a problem of hydrodynamics rather than of gravitation, as far as the baryonic component of the galaxy and small scales are concerned.

The idea discussed in this work is that the fragmentation of a cold medium is first controlled by the supersonic turbulent velocity field, while gravity takes over only intermittently. The large scale velocity field is determined by the gravitational potential, but on scales much smaller than the size of the galaxy the flow arises from hydrodynamical interactions inside the large scale flow; these interactions appear as energy cascades (the Reynolds number is very large) that are able to feed turbulent velocity fluctuations on small scales.

The velocity fluctuations are supersonic so that the fragmentation of the density field has little to do with gravitational forces, at this stage. In this way a very clumpy density distribution arises out of a complex system of shock waves; it is on such a distribution that gravity applies, forcing the largest or densest clumps to fragment into protostars.

If the first stage of the fragmentation based on turbulent supersonic cascades produces only very small clumps, as in the cold Galactic ISM (fig. 4), most of these will be stable against the gravitational collapse. In this case it is the intermittency in the statistics of the density field (shaped by supersonic turbulence) that determines what fraction of the total mass becomes unstable and is converted into protostars.

Therefore turbulence could be the reason why present day star formation is observed to be a slow process, capable of converting only a few percent of the gaseous mass into protostars (Duerr, Imhoff and Lada 1982; Myers et al. 1986; Mooney and Solomon 1988) in a period of about $10^7 years$ (Blitz and Shu 1980) after which the stellar feedback destroys the star formation site (Herbig 1962; Field and Saslaw 1965; Elmegreen and Lada 1977; Whitworth 1979; Lada and Gautier 1982). In fact the formula (27) for the protostar formation efficiency derived here predicts the conversion into protostars of only a few percent of the total mass after $10^7$ years, if applied to Galactic molecular clouds. Moreover the predicted mass distribution can be close to the observed Miller-Scalo mass distribution (Miller and Scalo 1979).

The main achievement of the present work is that of showing that *supersonic turbulence makes gravitational fragmentation and thus star formation inefficient* with the result that *the protostar formation rate increases with increasing average temperature and decreasing velocity dispersion.*

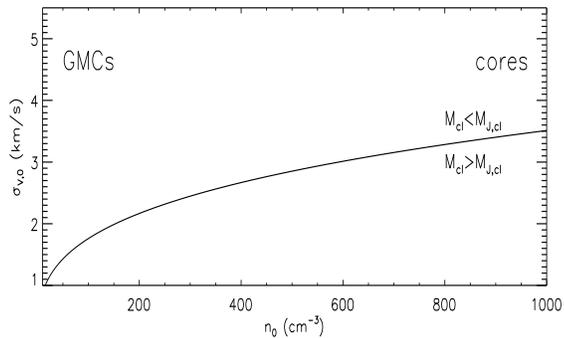

**Figure 4.** The turbulent fragmentation mass scale is compared with the Jeans' mass in the fragmented gas in the plane density-velocity dispersion. $M_{cl} = M_{J,cl}$ along the line; the region above the line produces turbulent fragments that are smaller than their Jeans' mass. The position of giant molecular clouds and their cores is shown to be in the stable part of the plane, meaning that gravitational fragmentation of the cold ISM is hindered by supersonic turbulence. The cores position is at the same velocity as the parent GMC because the velocity is referred to a fixed large scale (this is a consequence of Larson's relation).

Since the role of the magnetic field has not been considered one should be cautious when trying to predict present day star formation using the formulae here derived; nevertheless the present work shows that the turbulent fragmentation is certainly a powerful mechanism to slow down the process of star formation and even to give the stellar mass distribution its initial shape.

### 5.1 ISM Clumpiness

The present model relies on the basic idea that the ISM, due to ubiquitous supersonic turbulent flows, is fragmented down to the scale defined by equation (10), that marks the transition to subsonic motions. Therefore most of the mass is supposed to be contained in fragments of that typical scale, so that if a mass distribution is assumed its exponent has to be $\beta > 1$.

Most analysis of molecular cloud maps have revealed the presence of a very clumpy structure, with clumps mass distributed on power laws with exponent (defined in subsection 5.3) $0.0 < \beta < 0.7$ (Myers, Linke and Benson 1983; Blitz 1987; Carr 1987; Loren 1989a; Stutzkie and Güsten 1990; Lada, Bally and Stark 1991; Nozawa et al. 1991; Langer, Wilson and Anderson 1993; Williams and Blitz 1993).

This means that most of the detected CO mass is contained in the few largest clumps identified, with mass of the order of $100 - 1000 M_\odot$ and internal velocity dispersion larger than $1 km/sec$. These distributions are determined for clumps in the mass range $10 - 1000 M_\odot$, because of incomplete sampling for smaller clumps, and always detect cores with supersonic velocity dispersion.

Therefore almost none of the identified cores is a clump as defined in the present work (equation (10)), that is a fragment with nonsupersonic internal velocity dispersion: most of the observed cores must be internally fragmented. The predicted mass scale of turbulent fragmentation in the ISM is of the order of $0.1 M_\odot$, that is in fact two orders of magnitude smaller than the lower limit of most mass distributions which have been reported in the litterature.

Moreover it has been recognized that the ISM exhibits a hierarchical structure (Scalo 1985; Falgarone and Perault 1987; Vázquez-Semadeni 1994). This fact alone is sufficient to prove that the mass distributions obtained from cloud maps are not intrinsic distributions of fundamental fragments, but rather the result of the selection of some step in the hierarchy as the consequence of a combination of observational resolution, blending, projection and cores definition.

For these reasons the mass distributions obtained from the analysis of molecular cloud maps cannot be compared with the mass distribution used in the present work.

On the other hand high resolution observations (Walmsley et al. 1995; Velusamy, Kuiper and Langer, 1995; Langer et al., 1995) are proving that dark cloud cores are in fact fragmented down to very small scale into dense, unbound, young clumps with subsonic internal velocity dispersion, as expected from the present theoretical scenario. These clumps are transient structures not formed by gravity.

High resolution observations of very high density tracers are not available yet for a statistical analysis of a large number of such small clumps whose mass distribution has never been observationally determined.

### 5.2 Protostar formation

A sub-Jeans fragment of shocked gas is in general a transient structure in the ISM density field. Supersonic turbulence produces continuously such transient structures that are not able to collapse, as shown in fig. 4. This is a natural explanation, not invoking magnetic fields, of the fact that giant molecular clouds are not turning efficiently into stars in one free-fall time although they contain several Jeans' masses: supersonic turbulence makes gravitational instability behave as an intermittent process.

The protostar mass distribution is mainly determined by the p.d.f. of the turbulent density field that is expected to be close to lognormal. This fact explains why the stellar mass function can be much steeper than the mass distribution of cloud cores (see the previous subsection), although an alternative explanation is given in Zinnecker (1989). In any of the observed cores several stars can be formed with masses determined by the universal statistical properties of turbulence.

Several complications can be included for a better description of the formation of protostars and were not considered in this work.

First of all one Jeans' mass can fragment into a multiple stellar system instead of becoming a single protostar, and in this process only some fraction of the total mass will go into the final stars.

Secondly, gravitationally bound groups of transient clumps can be formed. Clumps in such groups can coalesce and form an unstable core that will collapse. Therefore when the present model predict no star formation, clumps coalescence could be the main star formation mechanism. Notice however that the collisional build up of an unstable core must be the consequence of the dissipative contraction of a bound system of smaller clumps, and not of collisions in general, which are just one aspect of the supersonic turbulent flows that are here described. Since an appreciable fraction of identified cores seem to be unbound systems (for example in Blitz 1987; Carr 1987; Loren 1989b; Herbertz, Ungerechts and Winnewisser 1991; Blitz 1993), the protostar formation via clump coalescence might be very inefficient. The largest cores are more likely to be gravitationally bound, but since their internal velocity dispersion is supersonic, turbulent fragmentation operates against protostar formation also inside these cores.

## 6 SUMMARY AND CONCLUSIONS

A supersonic turbulent flow has been used as a model for the velocity field in the cold ISM.

A typical mass of postshock clumps is defined and the stability of these to the gravitational collapse is studied.

The unstable postshock clumps are further fragmented into protostars by gravitational instability.

A turbulent spectrum and a mass and density distribution of the clumps are used to quantify the protostar mass spectrum and formation efficiency. It is shown how these depend on the ambient parameters.

The main aspects of the present statistical description of protostar formation can be summarized in the following points:

(i) the protostar mass distribution has a single maximum and depends on the ambient parameters.
(ii) The characteristic protostar has a mass $M_{J,cl} \propto n_0^{-1/2} T_0^2 \sigma_v^{-1}$, different from the traditional Jeans' mass and from the Jeans' mass with turbulent internal pressure, and equivalent to the Jeans' mass at constant external turbulent pressure.
(iii) The protostar formation efficiency is higher for larger mean density, larger temperature, lower turbulent velocity dispersion and, in the regime $M_{rad} > M_{cl}$, longer postshock cooling time (e.g. very low metallicity). The following power laws are followed: $e \propto n_0^{\frac{3}{2}(\beta-1)} T_0^{\beta-1} \sigma_{v,0}^{-5(\beta-1)} L_0^{3(\beta-1)}$, where $\beta > 1$ is the exponent of the clumps mass distribution.
(iv) the efficiency is quite sensitive to the ambient parameters and therefore to the dynamical evolution of the star forming system.

The mode of protostar formation here discussed is relevant to environments with supersonic turbulent velocity fields, postshock cooling time shorter than the large scale dynamical time and magnetic pressure not larger than thermal pressure. Whenever these conditions are met it is likely that the study of highly supersonic turbulence will give considerable insight into the problem of large scale protostar formation.

Problems such as the p.d.f. and the power spectrum of the density field, or the clumps density and mass distribution, and the role of the magnetic field will be tackled by detailed studies of three-dimensional supersonic MHD numerical turbulence (Nordlund & Padoan, in preparation).

It is worth noting that if the present mode of protostar formation predicts very low efficiencies it is likely that no other mode will be important because the medium is expected to be continuously disrupted into very small fragments (compared to some local Jeans' mass). The only restriction for the validity of the results here obtained is therefore that the magnetic pressure has to be smaller than the thermal pressure. Since the cosmological seed magnetic field is expected to be $\leq 10^{-17} G$ (Lazarian 1992), it can be argued that the present study is relevant to the problem of primordial star formation that occurs at epochs when no large scale dynamo has yet been activated in galaxies (Moffat 1978, Parker 1979, Krause et al. 1980, Ruzmaikin, Shukurov and Sokoloff 1988).

## ACKNOWLEDGEMENTS

This paper could not have been written without the guidance of my present supervisor Bernard J. T. Jones and of my previous supervisor prof. Guido Barbaro.

Bernard Pagel, Jesper Sommer-Larsen, Arthur D. Chernin gave useful advice.

Bernard J. T. Jones has carefully read and discussed the manuscript.

Comments by the referee B. G. Elmegreen were very helpful and are gratefully aknowledged.

The work has been supported in part by the Italian foundation "Ing. Aldo Gini" and by the University of Padova, during a visit at the Niels Bohr Institute in Copenhagen, and in part by the Danish National Research Foundation through its support for the establishment of the Theoretical Astrophysics Center.

## APPENDIX: List of symbols

| Symbol | Description |
| --- | --- |
| $\beta$ | exponent of the clumps mass distribution |
| $c_s$ | isothermal ambient sound speed |
| $e$ | protostar formation efficiency |
| $L_0$ | size of the region with r.m.s. velocity $\sigma_{v,0}$ |
| $m$ | protostar mass relative to $M_{J,cl}$ |
| $m_{cl}$ | clump mass relative to $M_{cl}$ |
| $M_{cl}$ | mass of the typical postshock clump |
| $M_{J,cl}$ | Jeans' mass in the typical clump |
| $M_J$ | local Jeans' mass |
| $M_{rad}$ | minimum mass scale for isothermal shocks |
| $M_*$ | total mass in protostars |
| $M_{tot}$ | total mass |
| $\mathcal{M}$ | Mach number |
| $\mathcal{M}_t$ | Mach number of the ambient r.m.s. turbulent velocity |
| $\mathcal{M}(100pc)$ | Mach number on the scale of $100pc$ |

| Symbol | Description |
|---|---|
| $n$ | local number density |
| $n_0$ | ambient average number density |
| $n_{cl}$ | number density of the typical fragments of mass $M_{cl}$ |
| $P(m)$ | Jeans' mass distribution |
| $P(x)$ | density fluctuations distribution |
| $P'(m)$ | protostar mass distribution |
| $P_{cl}(m_{cl})$ | clumps mass distribution |
| $\sigma_v$ | r.m.s. turbulent velocity |
| $\sigma_{v,0}$ | ambient r.m.s. turbulent velocity |
| $\rho$ | volume density |
| $\rho_0$ | ambient mean volume density |
| $r_{tr}$ | scale of the transition to subsonic turbulent motions |
| $T_0$ | ambient average temperature |
| $\tau_{cool}$ | postshock gas cooling time |
| $\tau_{dyn}(r)$ | dynamical time on the scale $r$ |
| $\tau_{dyn,0}$ | dynamical time on the scale $L_0$ |
| $x$ | local number density relative to $n_0$ |